\documentclass[prd,twocolumn]{revtex4}

\usepackage{amsmath}    
\usepackage{graphicx}   
\usepackage{verbatim}   
\usepackage{color}      
\usepackage{subfigure}  
\usepackage{hyperref}   
\usepackage{acronym}

\newcommand{\kahler}{K\"{a}hler }
\def\Tr{\mathop{\rm Tr}}

\begin{document}

\title{Stimulated Supersymmetry Breaking}

\author{Matthew McCullough}
\email{mccull@thphys.ox.ac.uk}
\affiliation{Rudolf Peierls Centre for Theoretical Physics, University of Oxford, 1 Keble Road, Oxford, OX1 3NP, UK}

\date{\today}

\begin{abstract}
We show that small soft terms can create a supersymmetry breaking minimum along a pseudo-flat direction of a hidden sector which would otherwise be incapable of spontaneous supersymmetry breaking.  As this minimum lies along a pseudo-flat direction, with non-zero F- or D-terms, the resultant supersymmetry breaking in the hidden sector can be orders of magnitude greater than the soft terms which created the minimum.  This opens new avenues for building models of gauge mediation, may have consequences for the cosmology of hidden sectors, and strengthens the case for multiple supersymmetry breaking sectors.
\end{abstract}

\preprint{OUTP-10-29P}

\maketitle

\section{Introduction}\label{intro}
If we wish to employ weak scale supersymmetry (SUSY) as a remedy to the hierarchy problem then, in addition to the supersymmetric standard model, a supplementary sector which spontaneously breaks SUSY becomes a necessity.

Extending this, if one entertains the idea of multiple sectors, with dynamics over a range of scales then, once supersymmetry has been broken in at least one sector, this raises the possibility of another hidden sector which may have soft masses well below the typical supersymmetric mass scales in the sector.  Such a sector would therefore be approximately supersymmetric; the soft supersymmetry breaking a small perturbation on the potential and dynamics of the sector.

In this work we are concerned with an effect peculiar to approximately supersymmetric hidden sectors, where, in analogy with approximate global symmetries, we define an `approximately' supersymmetric sector as one in which SUSY is broken softly at scales much lower than the supersymmetric mass scales in the sector.

One would expect such a small perturbation on the sector to have correspondingly small effects, however we comment here on a scenario whereby small soft parameters can have a pronounced effect, enabling a sector which has no stable SUSY breaking vacuum in the absence of the small soft terms to break SUSY spontaneously at typical mass scales of the sector, much greater than the SUSY breaking due to the soft terms.

The basic mechanism is as follows:  Consider a supersymmetic hidden sector, containing multiple superfields, which possesses a supersymmetric vacuum.  However, along the minimum of the scalar potential, far from the supersymmetric minimum, there also exists a tree-level flat direction with non-zero SUSY breaking.  Due to the SUSY breaking this flat direction is lifted radiatively, rendering it pseudo-flat, sloping gently towards the supersymmetric minimum.  Such a scalar potential is depicted in Figure \ref{secondary1}, and we call this sector the `secondary' sector.

Now we consider perturbing this secondary sector by adding a soft potential with scalar masses, $\tilde{m}$, much smaller than the typical supersymmetric mass scales in the secondary sector.  These soft terms could be generated due to supersymmetry breaking in another sector, the `primary' sector.  This soft potential only modifies the full scalar potential of the secondary sector in a minor way.  However, if the symmetry structure of the secondary sector superpotential is different from that of the soft potential, it is possible that the minimum of the supersymmetric scalar potential does not coincide with the minimum of the soft potential.

In this scenario, if the minimum of the soft potential coincides with a point along the pseudo-flat direction then, owing to the extremely flat potential, a local minimum can appear.  Thus the small parameters in the soft potential are important in the context of the secondary sector as it possesses a flat direction.

The consequences of a local minimum along a pseudo-flat direction now become important, as along such a direction the SUSY breaking is non-zero, and even more importantly this SUSY breaking occurs at scales much larger than the soft masses that induced the local minimum.  Thus we have `stimulated' supersymmetry breaking.

In this way approximately supersymmetric secondary sectors can exhibit a rather curious phenomenon, spontaneously breaking the approximate SUSY, even when the exactly supersymmetric counterpart has no SUSY breaking minimum.

This is amusing, but seems rather contrived - breaking supersymmetry in order to break supersymmetry.  However stimulated SUSY breaking may have a place in beyond-the-standard-model physics.

The first obvious application is in theories of gauge-mediated SUSY breaking, in particular in the framework of `ordinary gauge mediation' (OGM), which we now summarize.  For a recent general treatment of OGM see \cite{Cheung:2007es}, and a recent review of gauge mediation \cite{Kitano:2010fa}.  In this scenario the details of the hidden SUSY breaking sector are considered irrelevant, and the SUSY breaking in the hidden sector is parameterized by a singlet `spurion' superfield, $X$, with $\langle X \rangle = \theta^2 F$.  The singlet is then coupled to $N$ pairs of messenger superfields $Q$, $\overline{Q}$, transforming in the $\boldmath{5} \oplus \overline{\boldmath{5}}$ of $SU(5) \supset G_{SM}$.

The superpotential, including messenger masses and couplings to the SUSY breaking singlet, is then taken to be;
\begin{equation}
W_{OGM} = \lambda X \overline{Q} Q + M \overline{Q} Q ~~,
\label{OGM}
\end{equation}
and from this starting point SUSY breaking is communicated to the supersymmetric standard model (SSM) via gauge interactions, yielding a distinctive SSM spectrum.  Note that in Equation \ref{OGM} we have not taken the canonical assumption of $\langle X \rangle = M + \theta^2 F$ as the messenger masses could, in principle, arise from a different supermultiplet, especially when the scales of $M$ and $F$ are greatly separated.

We can see that the details of the hidden sector are necessarily non-trivial by considering the simplest model of SUSY breaking, a Polonyi model, with a hidden sector superpotential $W_h = f X$.  Coupling this to the messenger sector we have the total superpotential:
\begin{equation}
W_{T}= X (\lambda \overline{Q} Q + f) + M \overline{Q} Q ~~.
\label{Tot}
\end{equation}
Now, even if $\lambda$ is small, this theory possesses a supersymmetric minimum, and the previously flat $X$ directions are lifted by the messenger interactions.  These pseudo-flat directions tilt towards the supersymmetric minimum and the theory is rendered redundant for the purposes of SUSY breaking.

However, in the standard scenario the hidden sector involves couplings between $X$ and some extra superfields which generate radiative corrections to the potential for $X$, a long-lived SUSY breaking minimum can be achieved, and we can proceed with models of OGM.

With the aid of stimulated SUSY breaking we can proceed along an alternative avenue.  We start by positing the existence of an extra primary SUSY breaking sector which generates a soft mass $\tilde{m}$ for $X$, satisfying $\tilde{m}^2 < f \lesssim M^2$ \footnote{In fact $\tilde{m}^2$ can only be a loop factor smaller than $f$ for this mechanism to work, as we discuss in Section \ref{examp}.}.  This soft mass can stabilize $X$ along a SUSY breaking pseudo-flat direction, possibly near the origin, with $\langle X \rangle = \theta^2 f$.

The extra primary SUSY breaking sector need only break SUSY spontaneously, and is not lumbered with any of the necessary properties of a sector required for gauge mediation.  The strongest such requirement being that of significant R-symmetry breaking, along with SUSY breaking, in order to generate comparable gaugino and sfermion masses.

Hence a new recipe for building models of gauge mediation is to construct a secondary sector, including messengers, which possesses a tree-level flat direction along which there is comparable SUSY and R-symmetry breaking.  Even if this sector is incapable of spontaneous SUSY breaking in isolation, if we introduce small soft masses for the fields in this sector it may be possible to create a SUSY breaking minimum somewhere along the flat direction.  These small soft masses come at the price of a supplementary primary SUSY breaking sector, however this sector needn't also break R-symmetry.  Further, this primary SUSY breaking sector can be very weakly coupled to the stimulated SUSY breaking sector.  Finally the stimulated SUSY breaking in the secondary sector can be gauge-mediated to the SSM.

A novel feature of this set-up is that the gravitino mass is no longer set by the SUSY breaking in the sector which contributes dominantly to SUSY breaking in the SSM.

The mechanism of stimulated SUSY breaking also has some more general consequences, in particular in the context of multiple SUSY breaking sectors, whose phenomenology has been a subject of recent interest \cite{Benakli:2007zza,Cheung:2010mc,Cheung:2010qf,Craig:2010yf}.  If there exist multiple sequestered sectors then some of these may have been stimulated into breaking SUSY, even if they are incapable of spontaneous SUSY breaking in isolation.  Or, in other words, if there exist multiple SUSY breaking sectors then many of them needn't take the form of a successful SUSY breaking sector in isolation.

Another consequence follows from the observation that the gravitino mass, and thus soft SUSY breaking parameters, are of the order of the Hubble parameter, i.e.\ $m_{3/2} \sim H$, in the early Universe \cite{Dine:1995uk}.  It therefore follows that due to larger soft-masses, additional sectors may have experienced stimulated SUSY breaking in the early Universe, and, when $m_{3/2}$ fell below a critical value, such sectors could have experienced a SUSY-restoring phase transition.

In this work we focus mainly on the mechanism of stimulated SUSY breaking and comment briefly on the model building applications and cosmological consequences.  In Section \ref{examp} we construct both F-term and D-term examples of stimulated SUSY breaking.  We calculate the details of the F-term model for specific parameters, explicitly demonstrating stimulated SUSY breaking.  In Section \ref{mech} we discuss some general features of the mechanism, including the required properties of a candidate stimulated SUSY breaking sector.  Finally, we conclude in Section \ref{disc}.

\section{Some Illustrative Examples}\label{examp}
Before discussing the mechanism of stimulated SUSY breaking in more generality in Section \ref{mech}, we first focus on specific examples, in order to illustrate the mechanism as clearly as possible.

\subsection{An F-term Model}\label{F-termmod}
The model contains two sectors;
\begin{itemize}
\item  The `primary' sector, which breaks SUSY spontaneously and has dimensionful parameters of order $\Lambda_p$.  We denote the primary SUSY breaking superfield $P$.
\item  The `secondary' sector, which experiences stimulated SUSY breaking.  We denote the secondary SUSY breaking superfield $S$.   The dimensionful parameters in the secondary sector are of order $\Lambda_s$, and satisfy $\Lambda_s \ll \Lambda_p$.  In isolation this secondary sector does not possess a local SUSY breaking minimum, and relies on couplings to the primary sector in order to achieve stimulated SUSY breaking.
\end{itemize}

In this model we work in the limit of rigid SUSY, with $M_P \rightarrow \infty$.  We also assume a minimal \kahler potential for all fields, with the exception of the interaction term, which we introduce later.

We denote the superpotential for the primary sector $W_p$, and for the secondary sector $W_s$.  Thus the complete superpotential is $W = W_p + W_s$, where;
\begin{eqnarray}
W_p & = & P (y_p \phi_1 \phi_2 - h_p \Lambda_p^2) - \Lambda_p (\phi_1 \phi_3 + \phi_2 \phi_4)\\
W_s & = & S (y_s \overline{Q} Q - h_s \Lambda_s^2) - \Lambda_s \overline{Q} Q ~~. 
\label{fullW}\end{eqnarray}

We now describe the origin, and motivation, for each sector.

$W_p$ takes the form of the low energy limit of a well-known model of dynamical SUSY breaking, namely the Intriligator-Seiberg-Shih models \cite{Intriligator:2006dd}.  In \cite{Intriligator:2006dd} it was shown that SUSY QCD, with $N_f$ massive quark flavours, $N_f$ being in the range $N_c < N_f < 3/2 N_c$, exhibits a metastable, SUSY breaking vacuum if the quark masses are hierarchically smaller than the IR strong-coupling scale of the theory, i.e.\ $m_f \ll \Lambda_{IR}$.  This theory can be studied using the Seiberg dual \cite{Seiberg:1994pq} theory with $N_f$ flavours and $\tilde{N} = N_f - N_c$ colours.  Discarding supplementary massive, SUSY-preserving superfields, at low energies one is left with an O'Raifeartaigh-like superpotential of the form in $W_p$.  In this case $W_p$ is a simplified version of the model, and realistic scenarios with $\tilde{N} > 1$ contain $\tilde{N}$ copies of the dual quark superfields $\phi$.  Radiative corrections stabilize $P$ at the origin, and the dimensionful scales $\Lambda_p$ and $h_p \Lambda_p^2$ can be generated dynamically.  For our purposes $W_p$ is a well-motivated spontaneous SUSY breaking model, with dynamically generated scales \cite{Witten:1982df}, and for more details on the ISS models we refer the reader to the original work \cite{Intriligator:2006dd}.

$W_s$ is inspired by the form of the OGM set-up, and by early models of hybrid supersymmetric inflation \cite{Copeland:1994vg,Dvali:1994ms,Ross:1995dq,Craig:2008tv}.  It does not respect an R-symmetry, and is incapable of spontaneous SUSY breaking on its own.  $S$ is a gauge-singlet and we identify the standard model messengers, $Q$ and $\overline{Q}$, with a $5$ and $\overline{5}$ of $SU(5)$, with the intention of gauge-mediating to the supersymmetric standard-model (SSM).

This superpotential is not generic, however we can justify this form with selection rules from the spontaneous breaking of an R-symmetry due to gaugino condensation, as well as attempting to generate the parameters dynamically, along the lines of e.g.\ \cite{Dine:2006gm,Aharony:2006my}.  We introduce a pure SUSY QCD sector which becomes strongly coupled at a scale $\Lambda$.  The gauge fields are contained in the chiral superfield $W_\alpha$ and the massive quark superfields in this sector are $\overline{q}$, $q$.  If we allow non-renormalizable couplings of this sector to the messengers through physics at some higher scale $M_{\star}$, we can replace $W_s$ with an R-symmetric superpotential:
\begin{equation}
W_{s_R} =  \left (\frac{\overline{Q} Q}{M_\star} - m_q \right ) \overline{q} q + \Tr (W_\alpha^2) \left(\frac{1}{g^2_{SYM}} - \frac{\overline{Q} Q}{M_\star^2} \right)
\end{equation}
Thus if the extra SUSY Yang-Mills sector exhibits gaugino condensation we can identify $\langle \Tr (W_\alpha^2) \rangle \sim \Lambda^3$.  Further, below the strong coupling scale $\Lambda$ we can treat $\overline{q} q$ as a meson superfield, making the identification $\overline{q} q \sim \Lambda S$.  Hence we arrive at the form of $W_s$ given in Equation \ref{fullW}, where the parameters can be retrofitted as:
\begin{eqnarray}
y_s & \sim & \Lambda / M_\star \\ 
h_s \Lambda_s^2 & \sim & m_q \Lambda \\
\Lambda_s & \sim & \Lambda^3 / M_\star^2 ~~.
\end{eqnarray}
We also assume that the quark masses $m_q$ can be similarly retrofitted.  Although we have discussed a possibility for explaining the structure of $W_s$ we note that non-generic superpotentials can be maintained from high scales due to the perturbative non-renormalization of superpotentials.  We have also attempted to generate scales in this model dynamically, however one may alternatively choose to remain agnostic about the origin of the various terms, as the purpose of this model is mainly demonstrative.

For interactions between each sector we assume a \kahler potential term of the form;
\begin{equation}
K_{ps}  = -\frac{\lambda^2}{M_M^2} (P^{\dagger} P) (S^{\dagger} S)  ~~.
\label{Kint}\end{equation}
$K_{ps}$ is the only interaction between the primary and secondary sectors, in particular between the stimulated SUSY breaking superfield $S$ and the primary SUSY breaking superfield $P$.  We assume that $K_{ps}$ arises after integrating out messenger interactions between the primary and secondary sectors.  For the specific case of gravity mediation $M_M \sim M_{P}$, although lower mass messengers are also possible.

This completes the model.  Although the complete model does not exhibit an R-symmetry, it still breaks SUSY spontaneously and has no SUSY-preserving vacuum, even with kinetic mixing between $P$ and $S$ taken into account \footnote{We thank Graham Ross for bringing this to our attention.}.  At first this appears at odds with the general theorem of Nelson and Seiberg \cite{Nelson:1993nf}, however this is not the case, as although there is no continuous R-symmetry the superpotential is not generic and the primary and secondary sectors are not coupled through superpotential terms. In fact, the primary sector still breaks SUSY in the same vacuum that would be realized if the primary sector were isolated.

\subsection{Sectors in Isolation, $\lambda=0$}
To build a complete picture of stimulated SUSY breaking methodically it is necessary to first consider the sectors in isolation.

The primary sector, $W_p$, is an O'Raifeartaigh model, and with $h_p y_p < 2$ the minimum of the tree-level potential lies at $\langle \phi_i \rangle = 0$, with $\langle P \rangle$ undetermined, and a non-zero F-term of $F_P = h_p \Lambda_p^2$.  Although there exists a flat direction along $P$ at tree-level this is lifted radiatively.  Integrating out the massive $\phi_i$ superfields to find the Coleman-Weinberg potential \cite{Coleman:1973jx} one finds a one-loop soft mass for $P$.  Expanded to second order in the small parameter $h_p$ this is:
\begin{equation}
\tilde{V}_p = \frac{h_p^2 y_p^4}{24 \pi^2} \Lambda_p^2 |P|^2 ~~.
\label{softp}
\end{equation}
Thus $P$ is stabilized at the origin, and the low energy effective theory for the primary sector is given by Equation \ref{softp} and the superpotential:
\begin{equation}
{W_p}_{\text{eff}} = h_p \Lambda_p^2 P ~~.
\end{equation}

The story for the secondary sector, $W_s$, is somewhat different.  There exists a supersymmetric minimum at $\langle y_s S \rangle = \Lambda_s$ and $\langle \overline{Q} Q \rangle = h_s \Lambda_s^2/y_s$, and the vacuum behaviour away from this minimum is interesting.

Writing $\tilde{S} = y_s S -\Lambda_s$, which is zero at the supersymmetric minimum, then for $|\tilde{S}|^2 > y_s h_s \Lambda_s^2$ the scalar potential is minimized with $\langle \overline{Q} Q \rangle = 0$ and $\langle \tilde{S} \rangle$ is a flat direction at tree-level.  However in this region $F_S \neq 0$ and radiative corrections lift this flat direction.  Including these corrections by integrating out the massive $\overline{Q}$ and $Q$ superfields, the potential for $\tilde{S}$ is;
\begin{eqnarray}
V_s & = & h_s^2 \Lambda_s^4 + \frac{1}{32 \pi^2} (2 y_s^2 h_s^2 \Lambda_s^4 \log(|\tilde{S}|^2/\mu^2)) \nonumber\\
& + & (|\tilde{S}|^2 + y_s h_s \Lambda_s^2)^2 \log(1+y_s h_s \Lambda_s^2/|\tilde{S}|^2) \nonumber\\
& + & (|\tilde{S}|^2 - y_s h_s \Lambda_s^2)^2 \log(1-y_s h_s \Lambda_s^2/|\tilde{S}|^2)) ~~,
\label{loop}
\end{eqnarray}
where $\mu$ is the UV cut-off and we have written $\tilde{S} = y_s S -\Lambda_s$.

Whenever $|\tilde{S}|^2 \leq y_s h_s \Lambda_s^2$ the potential is minimized with $\langle \overline{Q} Q \rangle = (y_s h_s \Lambda_s^2 - |\tilde{S}|^2)/y_s^2$ and there are similar one-loop corrections.  These corrections are, however, subdominant in this region as $\tilde{S}$ becomes massive at tree level.  Although we don't give these corrections here they are included in all Figures.

\begin{figure}[t]
\centering
\includegraphics[height=2.9in]{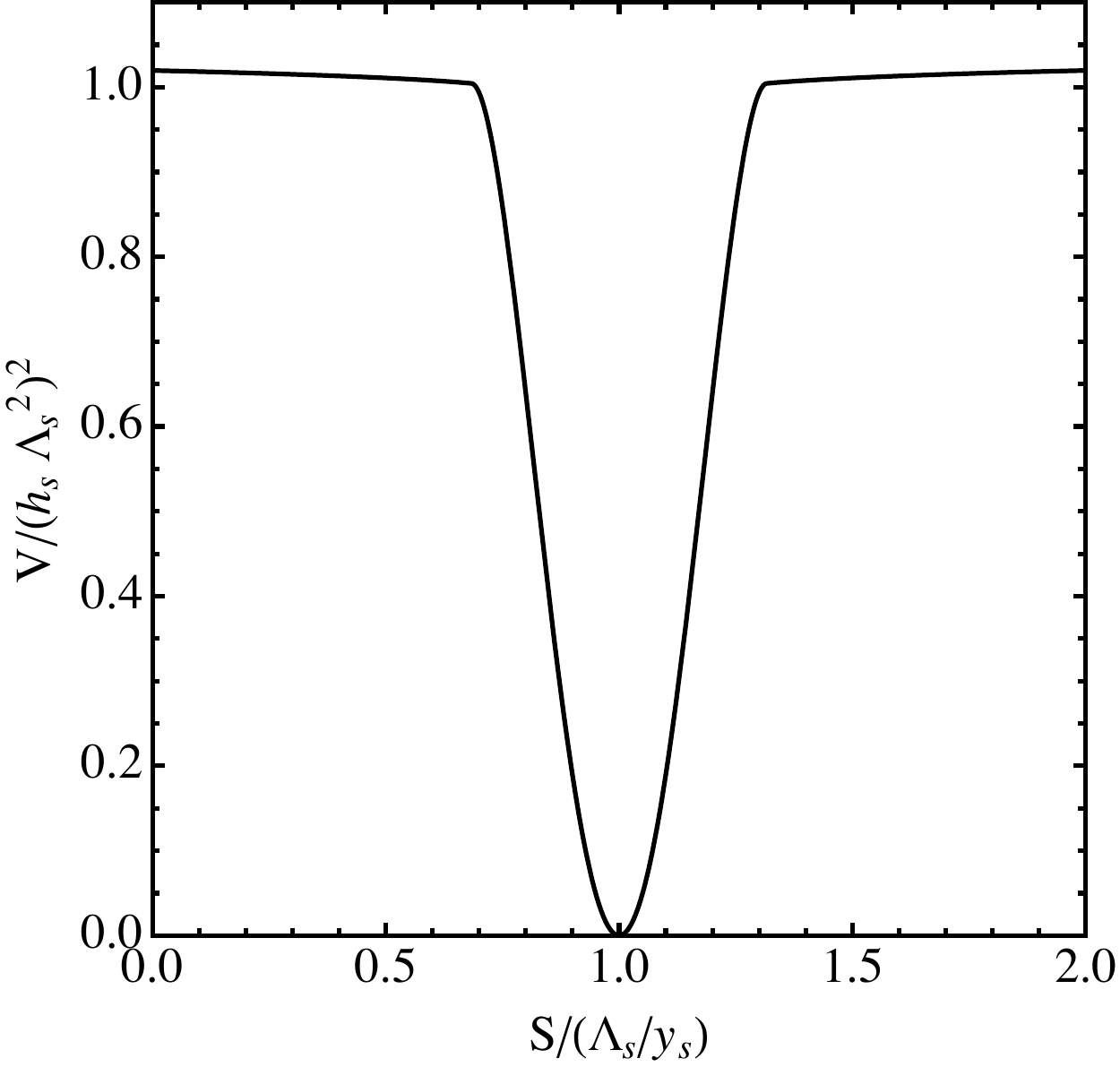}
\includegraphics[height=2.9in]{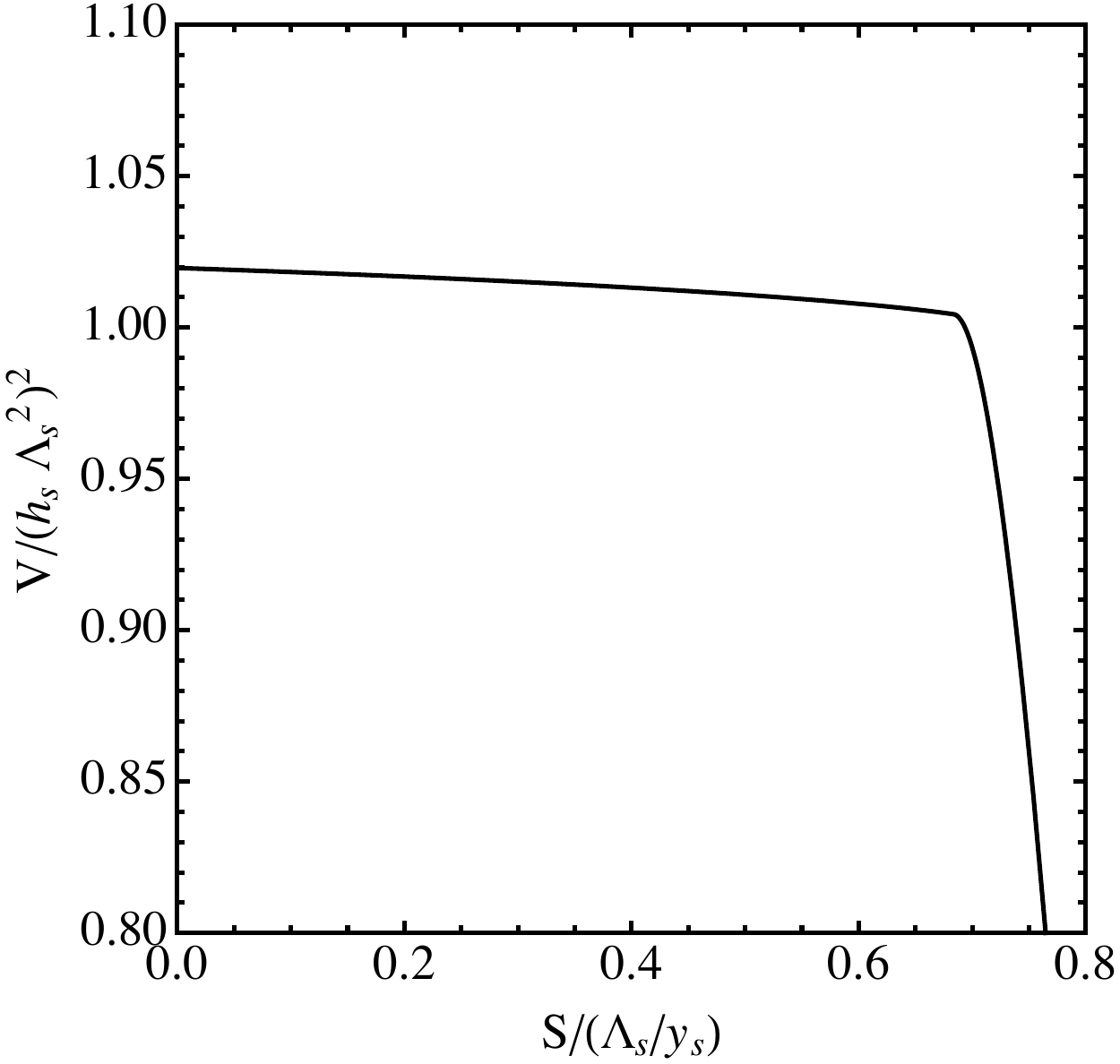}
\caption{The scalar potential for the secondary sector in isolation, i.e.\ in the limit $\lambda \rightarrow 0$.  The potential has been minimized in the $\overline{Q}$ and $Q$ directions, and is plotted as a function of $S$.  The relevant parameters are $h_s = 0.1$, $y_s = 1$, $\Lambda_s = 1$.  In the upper panel one can see the pseudo-flat directions either side of the supersymmetric minimum.  In the lower panel we see the lifting of these pseudo-flat directions due to radiative corrections.}
\label{secondary1}
\end{figure}

In the upper panel of Figure \ref{secondary1} we show the scalar potential for $S$, having minimized in the $\overline{Q}$ and $Q$ directions.  One can see the supersymmetric minimum and in the lower panel of Figure \ref{secondary1} we focus on the pseudo-flat direction to show the gentle slope generated by the radiative corrections.

This concludes the study of the primary and secondary sectors in isolation.  We now go on to consider the case where $\lambda \neq0$ and the SUSY breaking from the primary sector is mediated to the secondary sector.

\subsection{Stimulated SUSY breaking, $\lambda \neq0$}
Once we turn on the coupling, $\lambda$, between the primary and secondary sectors the SUSY breaking from the primary sector is mediated to the secondary sector.  In this case a soft mass for the scalar component of $S$ is generated, and, to second order in $h_p$, the soft potential for $S$ is:
\begin{equation}
\tilde{V}_S = \tilde{m}^2 |S|^2 = \frac{\lambda^2 h_p^2 \Lambda^4_p}{M_M^2} |S|^2  ~~.
\label{softs}
\end{equation}
This soft mass works to stabilize $S$ at the origin.  On the contrary we see from the lower panel of Figure \ref{secondary1} that the potential from the secondary sector alone leads to a shallow gradient at the origin, preferring to stabilize $S$ at the supersymmetric minimum $\langle y_s S \rangle = \Lambda_s$.  It is the interplay between the behaviour of these two terms that can lead to stimulated SUSY breaking.

In order to introduce a minimum near the origin the soft mass must be great enough to overcome the slope of the pseudo-flat direction.  In this model this corresponds to the requirement that:
\begin{equation}
\tilde{m} \gtrsim \frac{h_s y_s^2}{4 \pi} \Lambda_s ~~.
\end{equation}
Thus for relatively natural values of $h_s \sim y_s \sim 0.1$ the soft mass can be up to four orders of magnitude below the typical mass-scales in the secondary sector.  Smaller soft masses can lead to stimulated SUSY breaking if the couplings in the secondary sector are reduced.

\begin{figure}[t]
\centering
\includegraphics[height=2.9in]{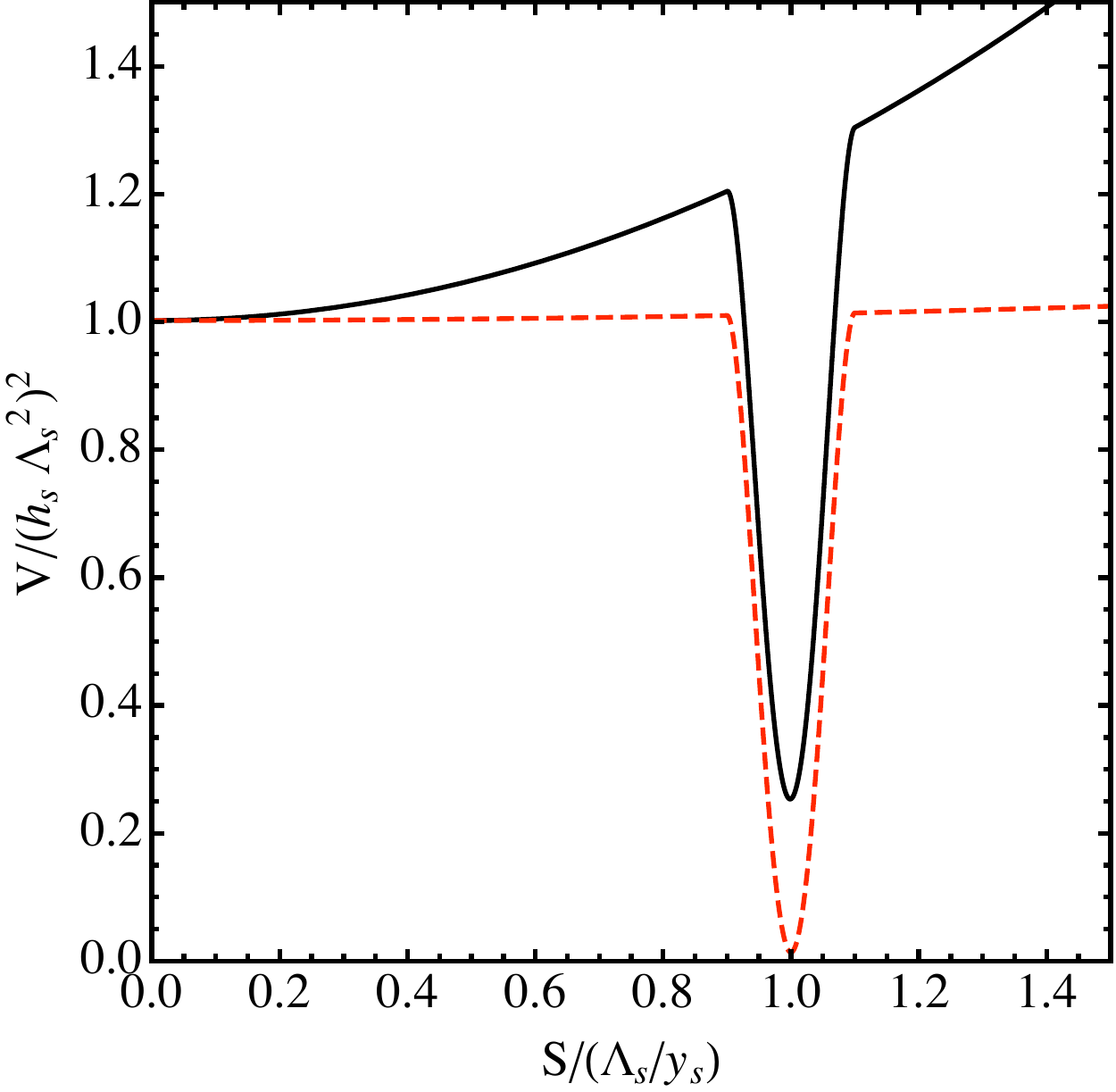}
\caption{The potential in the secondary sector, minimized in the $\overline{Q}$ and $Q$ direction, as a function of the scalar field $S$.  We have included the soft potential for $S$, generated through interactions with the primary sector, with $\lambda \neq 0$.  One can see that the soft mass has introduced a metastable minimum near the origin which would not exist otherwise.  The parameters chosen are $\Lambda_s = 10^4 \text{ TeV}$, $h_s = 0.1$, $y_s = 0.1$ and $\tilde{m} = 50 \text{ TeV}$ (solid black), $\tilde{m} = 10 \text{ TeV}$ (dashed red).  If we identify $\overline{Q}$ and $Q$ as messenger superfields, the metastable minimum gives the phenomenologically appealing value of $F_Q/M_Q = 100 \text{ TeV}$, where $F_Q$ is the SUSY breaking experienced by the messengers and $M_Q$ is the messenger mass.  Thus acceptable gaugino and sfermion masses can be generated from such a sector.}
\label{stimulatedfig}
\end{figure}
In Figure \ref{stimulatedfig} we plot the scalar potential for $S$ with the soft mass included.  One can see that a metastable minimum appears near the origin.  In this minimum the ratio of the soft masses to the stimulated SUSY breaking is, for the parameters we have chosen;
\begin{eqnarray}
F_{soft}/F_S & = & \frac{\tilde{m}^2}{h_s \Lambda_s^2} \nonumber\\
& = & 2.5 \times 10^{-4} ~~,
\end{eqnarray}
where the parameters used are detailed in the caption of Figure \ref{stimulatedfig}.  Thus, the soft terms have `stimulated' SUSY breaking in the secondary sector.

The soft mass of $\tilde{m} = 50 \text{ TeV}$ can be achieved with couplings of order $h_p \sim \lambda \sim 0.1$, messenger masses of $M \sim 10^9 \text{ TeV}$, and a primary SUSY breaking scale of $\Lambda_p \sim 10^6 \text{ TeV}$, below the limit at which one expects to generate dangerous flavour-changing neutral-currents from gravity mediated soft-terms.

In the scenario where the stimulated SUSY breaking is metastable one would like to make this vacuum cosmologically long-lived.  Approximating the potential with a triangle, the bounce action is;
\begin{equation}
B \approx \frac{2 \pi^2}{3} \left( \frac{\Lambda_s}{y_s \tilde{m}} \right)^2 \gg 1 ~~,
\end{equation}
where we have utilized the results of \cite{Duncan:1992ai}.  Thus the metastable SUSY breaking state can be made cosmologically long-lived.

Therefore, with $\overline{Q}$ and $Q$ identified as messengers charged under the standard model gauge group, one could mediate the stimulated SUSY breaking to the SSM, generating gaugino and sfermion masses at the same order of magnitude as a result of the comparable R-symmetry breaking and SUSY breaking scales in this model.  In such a manner, as well as being an amusing theoretical novelty, stimulated SUSY breaking can be utilized in building models of low-scale gauge mediation.

\subsection{A D-term Model}
It is also possible to build models of stimulated SUSY breaking where the SUSY breaking arises from D-terms.  A simple example appears in an adapted form of a D-term inflation model \cite{Binetruy:1996xj}.  We assume the same primary and interaction sectors as in Equation \ref{fullW}, and simplify the secondary sector superpotential to;
\begin{equation}
W_s = y_s S \overline{Q} Q - \Lambda_s \overline{Q} Q ~~,
\label{DW}
\end{equation}
which is invariant under an R-symmetry.  Further we introduce charges of $\pm 1$ for $Q$ and $\overline{Q}$ under an extra $U(1)_s$ gauge symmetry.  If we include a Fayet-Iliopoulos D-term, $h_s \Lambda_s^2$, the full tree-level scalar potential is:
\begin{eqnarray}
V_s & = & |y_s \overline{Q} Q|^2 + |y_s S - \Lambda_s|^2 (|Q|^2+|\overline{Q}|^2) \nonumber\\
& & + \frac{g^2}{2} (|Q|^2-|\overline{Q}|^2+h_s \Lambda_s^2)^2 ~~.
\label{DV}
\end{eqnarray}
Both dimensionful parameters in this model can also be retrofitted to dynamically generated scales, following the examples in \cite{Dine:2006gm}.  

The scalar potential in Equation \ref{DV} has a similar behaviour to the F-term model described previously.  There is a supersymmetric minimum at $\langle y_s S \rangle = \Lambda_s$, $\langle Q \rangle =0 $, $\langle |\overline{Q}|^2 \rangle = h_s \Lambda_s^2$, and for large values of $S$, when $|y_s S - \Lambda_s|^2>h_s g^2 \Lambda_s^2$, the scalar potential is minimized for $\langle Q \rangle = \langle \overline{Q} \rangle = 0$ and is tree-level flat in the $S$ direction.  Along this flat direction the SUSY breaking is:
\begin{equation}
D_s = g h_s \Lambda_s^2 ~~.
\end{equation}
Again, as SUSY is broken, this flat direction is lifted by radiative corrections.  For large $|\tilde{S}|^2 = |y_s S - \Lambda_s|^2 > h_s g^2 \Lambda_s^2$ the potential behaves as;
\begin{eqnarray}
V_s & = & \frac{g^2}{2} h_s^2 \Lambda_s^4 + \frac{1}{32 \pi^2} (2 g^4 h_s^2 \Lambda_s^4 \log(|\tilde{S}|^2/\mu^2)) \nonumber\\
& + & (|\tilde{S}|^2 + g^2 h_s \Lambda_s^2)^2 \log(1+g^2 h_s \Lambda_s^2/|\tilde{S}|^2) \nonumber\\
& + & (|\tilde{S}|^2 - g^2 h_s \Lambda_s^2)^2 \log(1-g^2 h_s \Lambda_s^2/|\tilde{S}|^2)) ~~,
\label{loopD}
\end{eqnarray}
where $\mu$ is a UV-cutoff.

Thus, once again, if we include the coupling of this secondary sector to the primary sector then soft terms for $S$, generated as a result of the SUSY breaking in the primary sector, can stabilize $S$ somewhere along this pseudo-flat direction.  In this case the overall SUSY breaking in the secondary sector is;
\begin{equation}
D_s = g h_s \Lambda_s^2 \gg \tilde{m}^2 ~~,
\end{equation}
and the SUSY breaking in the primary sector has stimulated SUSY breaking in the secondary sector.

Having considered some simple models of stimulated SUSY breaking we now go on to consider the mechanism in more general terms.

\section{Some General Features of Stimulated SUSY breaking}\label{mech}
Considering the mechanism in general, we can identify the key ingredients of a stimulated SUSY breaking set-up, as well as some generic properties.

First we consider the tree-level scalar potential of the secondary sector alone, and simply denote this potential $V$.  It is commonplace for supersymmetric theories to possess tree-level flat directions, where $V' = \partial V/ \partial \phi = 0$ and $V'' = \partial^2 V/ \partial \phi^2 = 0$, with $\phi$ a complex scalar field.  Usually such flat directions are supersymmetric, with $V_{\text{flat}} = 0$, however it is possible that flat directions arise which have non-zero F- or D-terms, giving;
\begin{equation}
V_{\text{flat}} = F_i^\dagger F_i + \frac{1}{2} D^a D^a \neq 0 ~~,
\end{equation}
where the sum over fields is implied.

Denoting the superfields corresponding to the scalar flat directions $X_i$, and denoting any additional superfields which are charged under gauge and global symmetries $\phi_j$, then a renormalizable theory of the following form may possess such flat directions;
\begin{equation}
W = f_i X_i + (\lambda_{ijk} X_i + M_{jk}) \phi_j \phi_k ~~,
\label{gen}
\end{equation}
where $\lambda_{ijk}$ is a trilinear coupling constant.  Depending on the nature of $\lambda_{ijk}$ and $M_{jk}$ it may be possible to justify the form of the superpotential with an R-symmetry, under which the singlet superfields have R-charges of $Q_{X_i} = 2$.  Otherwise, if the theory does not possess an R-symmetry, quadratic and cubic terms in $X_i$ can be forbidden with selection rules from spontaneous R-symmetry breaking \footnote{A further term, $A_{ijk} \phi_i \phi_j \phi_k$, may also be added.  However, depending on the form of $A_{ijk}$, this may allow for cancellations between terms which spoil the stimulated SUSY breaking.}.

Studying the F-terms from the superpotential in Equation \ref{gen} one can see that for very large field values, where $\sqrt{f_i} \langle \lambda_{ijk} X_i + M_{jk} \rangle \gg f_i$, the scalar potential is minimized with $\langle \phi_i \rangle = 0$, and is tree-level flat in the $\langle X_i \rangle$ direction.  Thus any theory of the form in Equation \ref{gen} may be a candidate for stimulated SUSY breaking, even if it possesses a SUSY-preserving minimum.

As the flat directions in such theories are not supersymmetric they are typically lifted radiatively, generating a potential for the flat direction.  If this is the case we call this direction `pseudo-flat'.  Therefore we now correct the tree-level potential with the full one-loop Coleman-Weinberg \cite{Coleman:1973jx} potential:
\begin{equation}
V_{F} = V + V_{l} ~~.
\end{equation}
Pseudo-flat directions are usually considered in the context of theories which break SUSY at tree-level.  In such a case, once radiative corrections are taken into account, a local minimum with non-zero vacuum energy exists, i.e.\ $V > 0$, $V'_F = 0$ and $V''_F > 0$, where only the tree level vacuum energy is included in the first inequality.  Attempting to stimulate SUSY breaking in a theory which already possesses a local SUSY breaking minimum would be rather pointless, therefore we focus our attention on theories which `almost' break SUSY.

We define an `almost' SUSY breaking theory as one in which there exists a supersymmetric minimum, but, far from the supersymmetric minimum, also exhibits a tree-level flat direction along the local minimum of the scalar potential, where the conditions;
\begin{eqnarray}
V > 0, & V' = 0, & V'' = 0 ~~,
\end{eqnarray}
are satisfied.  Although this looks promising for spontaneous SUSY breaking, in an almost-SUSY breaking theory the full potential, including radiative corrections, has no stable points where:
\begin{eqnarray}
V > 0, & V'_F = 0, & V''_F > 0 ~~.
\end{eqnarray}
Two examples of such a theory were discussed in the previous section and the upper panel of Figure \ref{secondary1} demonstrates such a potential.  In an almost-SUSY breaking theory, rather than stabilizing the fields at a local SUSY breaking minimum, the radiative corrections lift the flat direction so as to gently slope towards the supersymmetric minimum.

Now, if SUSY breaking occurs in the primary sector then soft masses, and in general a soft scalar-potential, can be generated in the secondary sector, along with supersymmetric superpotential terms generated via the Giudice-Masiero mechanism \cite{Giudice:1988yz}.  We assume that all such dimensionful terms are of order $\tilde{m} \sim F_P/M_M$, and coupling constants of order $c \sim \tilde{m}/M_M$, where $F_P$ parameterises the SUSY breaking in the primary sector and $M_M$ is the scale of the messengers between the primary and secondary sectors.

These terms will alter the scalar potential for the secondary sector and we call the additional terms in the scalar potential $\tilde{V}$, all terms vanishing in the limit $\tilde{m} \rightarrow 0$.  If $\tilde{m}$ is much smaller than the typical scales in the secondary sector then this additional soft scalar potential can be considered as a small deformation of the scalar potential, and the secondary sector becomes approximately supersymmetric.

In most cases such a small perturbation on the scalar potential is of little interest.  However, if $V$ contains a tree-level flat direction where $F \neq 0$, and $\tilde{V}$ possesses a stable minimum somewhere along this flat direction, it is possible for the scalar fields to be stabilized in a metastable, or even stable, state with SUSY breaking $F \gg \tilde{m}^2$.  This is `stimulated' SUSY breaking.

If a tree-level flat direction with non-zero F-terms were exactly flat at all orders then stimulated SUSY breaking could be achieved with arbitrarily small soft terms.  However, tree-level flat directions are typically lifted due to the SUSY breaking within the secondary sector.  Therefore in order to realize stimulated SUSY breaking in a realistic scenario the soft terms of order $\tilde{m}$ must overcome the gradient of the full one-loop potential.  This places a lower limit on the magnitude of SUSY breaking that must be transmitted to the secondary sector in order to stimulate SUSY breaking.  A rule-of-thumb is that soft terms should be greater than the typical scales in the secondary sector suppressed by a loop factor.

Finally we come to some general conditions for stimulated SUSY breaking.  If $V$ is the tree-level scalar potential for a secondary sector, $V_l$ is the one-loop correction to this potential, and $\tilde{V}$ is the soft scalar potential generated from SUSY breaking in the primary sector, then the full scalar potential is $V_T = V+V_l+\tilde{V}$.  To realize stimulated SUSY breaking we require that at some point in field space the conditions;
\begin{eqnarray}
\label{conditions}
V \neq 0, & V' = 0, & V'' = 0, \\
& \tilde{V}' + V'_l = 0 & \\
& \tilde{V}'' + V''_l > 0 &
\end{eqnarray}
are satisfied.  Thus, a stable local minimum of the soft-plus-loop-level scalar potential must coincide with a point along a SUSY breaking tree-level flat direction.

\subsection{A Pseudo-Goldstino}
Although stimulated SUSY breaking is spontaneous, the presence of the soft terms implies that it is only an approximate SUSY that is broken spontaneously.  Thus, in the rigid SUSY limit, we expect a pseudo-goldstino, as opposed to an exactly massless goldstino.  One can see that in the limit $\tilde{m} \rightarrow 0$, from Equation \ref{conditions} we still have $V \neq 0$ and $V' = 0$, and consequently a massless goldstino.  Therefore any correction to the mass must be of order $\tilde{m}$.  These corrections can arise at tree-level from SUSY-preserving superpotential terms which are generated via the Giudice-Masiero mechanism \cite{Giudice:1988yz}, or, if such terms are absent, then one would also expect a mass $m_{PG} < \tilde{m}$ to be generated at loop-level.

Therefore a `smoking gun' consequence of stimulated SUSY breaking would be the existence of a light fermion, with mass a few orders of magnitude below the scale of SUSY breaking in a hidden sector.  The mass of this pseudo-goldstino would give an estimate of the soft SUSY breaking terms which are stabilizing the stimulated SUSY breaking.  If $\tilde{m} \sim m_{3/2}$, corresponding to gravity mediation from the primary sector, one would expect the pseudo-goldstino to acquire mass $2 m_{3/2}$ \cite{Cheung:2010mc,Craig:2010yf}.  However, if the mediation from the primary to secondary sector occurred at lower scales, and $\tilde{m} \gg m_{3/2}$, the pseudo-goldstino mass would bear no relation to the gravitino mass.

\subsection{Non-minimal \kahler Potential}
Although in Section \ref{mech} the description of potential stimulated SUSY breaking scenarios was fairly general, in the individual sectors of Section \ref{examp} we assumed minimal \kahler potential terms.  However we also assumed that scales in each sector were generated dynamically, explaining the hierarchy between these scales and the Planck scale.  As one typically expects non-renormalizable operators when considering effective descriptions of strongly coupled theories it is necessary to briefly discuss the possible effects such operators could have on stimulated SUSY breaking.

Non-renormalizable operators in the primary sector are not a concern for this work, as all we require of the primary sector is that it breaks SUSY spontaneously and dynamically, and we are not concerned with the details of how this occurs here, although we note that higher order \kahler terms in the ISS model used in Section \ref{examp} are under control \cite{Intriligator:2006dd}.

The important consideration is whether higher-order terms, in particular \kahler terms, which can be consistent with imposed symmetries, may spoil the flatness of the pseudo-flat direction in the secondary sector.  A particularly troublesome operator is:
\begin{equation}
\Delta K_S = \frac{(S^\dagger S)^2}{\Lambda^2} ~~.
\end{equation}
If the unknown scale $\Lambda$ is of order the typical scales in the secondary sector, i.e.\ $\Lambda \sim \Lambda_s$, this operator generates a soft mass of order $\Lambda_s \gg \tilde{m}$ along the flat direction, thus spoiling stimulated SUSY breaking.  However, we expect that such an operator would arise from integrating out modes of mass $\Lambda_s$ and would come suppressed by at least a loop factor, so should not spoil the stimulated SUSY breaking.  It is in fact such corrections that we calculated in Equation \ref{loop}.

If $\Lambda \gg \Lambda_s$, arising from integrating out unknown heavier modes associated with some strong-coupling scale, we require that $\Lambda_s^2 \ll \tilde{m} \Lambda$ in order for stimulated SUSY breaking to be possible.

\subsection{Gravity Mediated Soft Terms}
A particularly appealing scenario for stimulated SUSY breaking would be if the primary sector were to break SUSY at scales $F_P \lesssim (10^{10} \text{ GeV})^2$, generating soft masses of order $\tilde{m} \sim 10 \text{ GeV}$ in the secondary, and SSM, sectors via gravity mediation.  These soft masses could in principle stimulate SUSY breaking in a hidden sector up to scales as high as $\sim 100 \text{ TeV}$, which would correspond to four orders of magnitude between the soft terms and the SUSY breaking.  

This gravity-mediated stimulated SUSY breaking can be realized with the secondary sector model of Section \ref{F-termmod}.  For $\tilde{m} \sim 10 \text{ GeV}$ we find that with the parameters $\Lambda_s = 10^6 \text{ TeV}$, $h_s = 1$ and $y_s = 10^{-4}$, a metastable SUSY breaking vacuum appears near the origin.  The small value of $y_s$ implies that the secondary SUSY breaking field and the messengers are very weakly coupled.  In addition, for these parameters, the ratio of SUSY breaking experienced by the messengers to the messenger masses is $F_Q/M_Q = 100 \text{ TeV}$, suitable for gauge mediation.  Summarizing this set-up; a primary SUSY breaking sector generates soft terms in all sectors via gravity mediation.  These soft terms go on to stimulate SUSY breaking in a secondary sector at a much higher scale, and this stimulated SUSY breaking is subsequently gauge mediated to SSM.

Such a scenario would also correspond to a neat realization of the `Goldstini' scenario \cite{Cheung:2010mc}, with both the gravitino, from the primary sector, and goldstino, from the secondary sector, with masses potentially measurable at the LHC, along with an appealing cosmology \cite{Cheung:2010qf}.  All arising from one fundamental SUSY breaking in the primary sector.

\section{Discussion}\label{disc}
We have demonstrated the mechanism of stimulated SUSY breaking, both in the context of specific models and in more general terms.  It may be that this mechanism is merely a technical novelty; allowing approximately supersymmetric sectors to break SUSY spontaneously, even though their exactly supersymmetric counterparts cannot.  However, if one is willing to pay the price of an extra SUSY breaking sector, stimulated SUSY breaking can open new possibilities for building models of gauge mediation, by utilizing sectors that, alone, are incapable of breaking SUSY spontaneously.

As a sector which experiences stimulated SUSY breaking need not exhibit an R-symmetry, it may be possible to avoid some of the problems faced in building models of gauge mediation which relate to the necessity of R-symmetry breaking.  One such difficulty lies in generating gaugino masses comparable to the sfermion masses.

Stimulated SUSY-breaking may also have interesting cosmological consequences as large soft masses, arising due to thermal effects in the early Universe \cite{Dine:1995uk}, may have led to stimulated SUSY breaking in sectors which have subsequently evolved to a SUSY-preserving minimum.  Further, if there exist multiple sequestered sectors, stimulated SUSY breaking lends support to the concept of multiple SUSY breaking sectors \cite{Benakli:2007zza,Cheung:2010mc}.  This is because, even if the extra sectors are not capable of spontaneous SUSY breaking in isolation, these sectors may have been stimulated into spontaneous SUSY breaking by the existence of a single original SUSY breaking sector.

Recently the apparently similar, but essentially different model of `cascade SUSY breaking' was proposed \cite{Ibe:2010jb}.  This model also relies on SUSY breaking from a primary sector which is mediated to a secondary sector, generating soft terms of order $\tilde{m}$.  The ultimate difference, however, is that in cascade SUSY breaking the eventual SUSY breaking in the secondary sector is $F_S \propto \tilde{m}^2$, corresponding to the global minimum in Figure \ref{stimulatedfig}.  Whereas, in stimulated SUSY breaking we have $F_S \propto \Lambda_s^2 \gg \tilde{m}^2$ and the soft terms stimulate the secondary sector to break SUSY at much higher scales.  Thus, although the overall set-up is rather similar, cascade, and stimulated SUSY breaking mechanisms both constitute appealing, although rather different, scenarios for model building.

\section{Acknowledgements}
I am grateful to James Barnard, Joe Conlon, Nathaniel Craig, Rhys Davies, John March-Russell and Graham Ross for stimulating conversations, and for comments on an early draft of this work.  I acknowledge support from an STFC Postgraduate Studentship and from University College, Oxford.

\bibliography{StimulatedSUSYrefs}

\end{document}